\documentclass[prl,superscriptaddress,amsmath,amssymb,floatfix,twocolumn,showpacs,amsfonts,longbibliography,prx]{revtex4-2}

\usepackage{graphicx}
\usepackage{dcolumn}
\usepackage{bm}

\usepackage{amsmath,amssymb}
\usepackage{graphicx}
\usepackage{xcolor}
\usepackage[colorlinks=true,linkcolor=blue,citecolor=blue,urlcolor=blue]{hyperref}
\usepackage{braket}
\begin{document}
	
	\title{Kinematic saturation of the electromagnetically-induced-transparency peak in thermal atomic vapors}
	\author{Sheng-Xian Xiao}
	\affiliation{Chongqing Key Laboratory for Strongly Coupled Physics, Chongqing University, Chongqing, 401331, China}
	\affiliation{Center for Quantum Materials and Devices, Chongqing University, Chongqing, 401331, China}
    
    \author{Jing-Yi Mao}
    \affiliation{Chongqing Key Laboratory for Strongly Coupled Physics, Chongqing University, Chongqing, 401331, China}
	\affiliation{Center for Quantum Materials and Devices, Chongqing University, Chongqing, 401331, China}
    
	\author{Tao Wang}
	\thanks{corresponding author: tauwaang@cqu.edu.cn}
	\affiliation{Chongqing Key Laboratory for Strongly Coupled Physics, Chongqing University, Chongqing, 401331, China}
	\affiliation{Center for Quantum Materials and Devices, Chongqing University, Chongqing, 401331, China}
	\affiliation{Center of Modern Physics, Institute for Smart City of Chongqing University in Liyang, Liyang 213300, China}
	\begin{abstract}
    We demonstrate that, in thermal atomic vapors, the electromagnetically‑induced‑transparency (EIT) transmission peak exhibits saturation. Investigating the amplitude of the velocity‑selected double‑resonance transmission peak---a purely kinematic effect arising from finite wave‑vector mismatch---we identify a threshold atomic column density $N_s$. For column densities $N$ below $N_s$, the amplitude of the transmission peak maximum increases with column density. Once it exceeds $N_s$, the peak amplitude maximum remains constant, while the corresponding detuning shifts away from resonance and scales as $\sqrt{\ln(N/N_s)}$. Furthermore, We find that the operating point for optimum photon‑shot‑noise‑limited sensitivity deviates from that of the transmission‑peak maximum. Our study provides guidance for optimizing EIT readout in vapor‑cell and for precision measurements built upon this scheme.
	\end{abstract}
	\maketitle

\textit{Introduction.}---Electromagnetically induced transparency (EIT) uses a control laser to create quantum interference that opens a narrow transparent window in an otherwise absorbing atomic medium~\cite{Harris1990,Boller1991,Fleischhauer2005}. Among all platforms, thermal atomic vapors, which require no cryogenics, are the workhorse of this physics: for counter-propagating lasers the Doppler shifts associated with the two transitions largely compensate one another, such that the interference persists despite Doppler broadening~\cite{GeaBanacloche1995,Xiao1995}. Vapor-cell EIT accordingly supports a broad range of applications, including slow light and quantum memories~\cite{Kash1999,Phillips2001,Lukin2003}, giant Kerr nonlinearities~\cite{Schmidt1996,Wang2001}, all-optical switching~\cite{Harris1998,Sheng2011}, precision magnetometry~\cite{Budker2002,Fabricant2023}, chip-scale frequency references~\cite{Kitching2018}, and SI-traceable microwave electrometry with Rydberg atoms~\cite{Sedlacek2012,Fan2015,Jing2020,Yuan2023}. For these applications, increasing the amplitude of the EIT transmission peak represents a straightforward way to boost measurement sensitivity and accuracy~\cite{Prajapati2021,Hu2025}. Previous studies on optimizations of the transmission peak~\cite{Su2022,Meyer2021} mainly focus on the EIT resonance condition (both lasers resonant with the respective atomic transitions), while recent experiments have moved beyond the resonance: tuning detunings to a selected thermal velocity class has enhanced the transmission by almost an order of magnitude~\cite{Hu2025}, which is the so called thermal resonance-enhanced transparency (TRET). This enhancement is considered to arise from that there are far more velocity‑selected fast atoms than slow atoms~\cite{Hu2025,cpl_42_12_120601}. In this letter, we show that this is not the case, identify what limits peak growth, and reveal how the detuning corresponding to the peak maximum evolves with atomic density.

\begin{figure*}[t]
\includegraphics[width=0.98\textwidth]{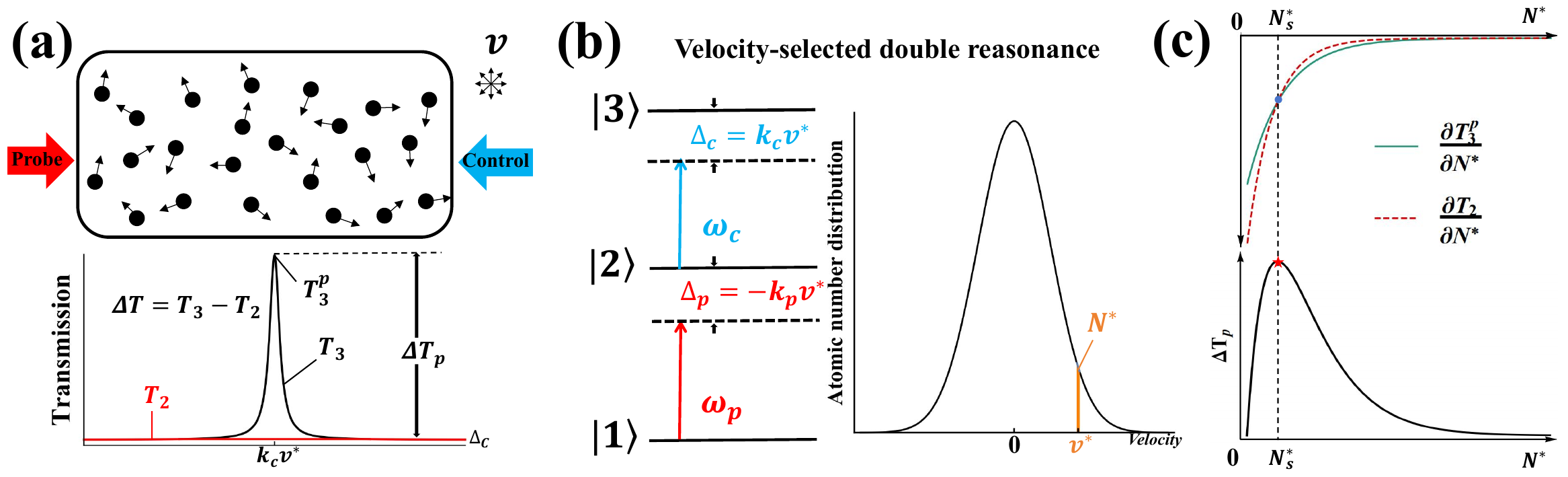}
\caption{\label{fig1}(a) EIT response of a thermal vapor driven by counterpropagating probe and control lasers. $T_3$ ($T_2$) is the probe transmission with (without) the control laser, $\Delta T \equiv T_3-T_2$, the transmission peak $T_3^p$ and peak contrast $\Delta T_p$ are attained at the control detuning $k_cv^*$ setting probe detuning at $-k_pv^*$. (b) Velocity-selected double resonance: for laser detunings $\Delta_p=-k_pv^*$ and $\Delta_c=k_cv^*$, the atoms of velocity class $v^*$ (shaded) are simultaneously on single- and two-photon resonance and contribute the participating column density $N^*$. (c) The rates at which $T_3^{p}$ (solid) and $T_2$ (dashed) vary with the participating column density $N^*$ cross at $N^*_s$, where the peak contrast $\Delta T_p$ is maximal; the crossing defines the optimal
participating column density.}
\end{figure*}

\textit{General picture.}---Consider a thermal vapor cell filled with ladder‑type three‑level atoms, which is driven by counter‑propagating probe and control lasers of distinct wave‑numbers [Fig.~\ref{fig1}(a)]. The transmission spectrum of probe laser is obtained by scanning the detuning of the control laser. The transmission contrast studied in this paper is defined as $\Delta T\equiv T_3-T_2$, where $T_3$ and $T_2$ denote the probe transmission with and without the control laser, respectively. In the large-detuning regime, $T_3=T_2$ since the atoms do not interact with the control laser. We next describe \emph{velocity-selected double-resonance} (VSDR) EIT~\cite{Hu2025}. Due to incomplete Doppler compensation, it is possible to set appropriate detunings for the control ($\Delta_c$) and probe lasers ($\Delta_p$) such that both fields are resonant solely with atoms within one specific longitudinal velocity group, as shown in Fig.~\ref{fig1}(b). The VSDR condition requires that $\Delta_c=k_cv^{*}$  and $\Delta_p=-k_pv^{*}$, where $k_c(k_p)$ is the wavenumber of control (probe) laser and $v^{*}$ is the selected atomic velocity. For fixed $\Delta_c$ and the corresponding $\Delta_p$, only a small fraction of atoms with velocities centered at $v^{*}=\Delta_c/k_c$ contributes to VSDR EIT transmission peak contrast $\Delta T_{p}=T_3^{p}-T_2$, in which $T_3^{p}$ is $T_3$ at the peak. The value of $\Delta T_{p}$ is therefore determined by the VSDR column density of atoms, denoted $N^*$, which represents the number of atoms per unit area along the probe beam path, see Fig.~\ref{fig1}(b). By investigating the dependence of \(\Delta T_{p}\) on \(N^*\), we find an optimal \(N^*\) (denoted \(N^*_\text{s}\)) for the VSDR EIT peak, given by $\partial T_3^{p} /\partial N^* =\partial T_2 /\partial N^*$, see Fig.~\ref{fig1}(c). The underlying physics is that $T_3^{p}$ and $T_2$ decrease at different rates with \(N^*\) increasing. 

Building on the above discussion, we now examine the relationship between the total column density $N$ and the maximum VSDR EIT peak contrast \(\Delta T_p^{m}\). Similarly, there exists a threshold $N_s$. For $N<N_s$, the value of \(\Delta T_p^{m}\) increases with increasing $N$, while the corresponding detunings of both lasers matches the EIT resonance. This is because the column density of atoms at zero longitudinal velocity is always smaller than $N^*_s$. When $N$ exceeds \(N_s\), the density of atoms at zero longitudinal velocity surpasses \(N^*_s\). Saturation of \(\Delta T_p^\text{m}\) is then expected, and the corresponding detuning moves away from resonance to a specific \(\Delta_c\), at which the column density of atoms with longitudinal velocity \(v^{*}=\Delta_c/k_c\) matches \(N^*_s\).  This is the physical explanation behind the TRET introduced in Ref.~\cite{Hu2025}, not the increasing of the number of atoms interacting with the laser. In the remainder of this paper, we demonstrate the above concept and discuss its implications for precision measurement within the photon-shot-noise limit.

\textit{Model.}---We consider a ladder three-level system in a vapor cell [Figs.~\ref{fig1}(a) and \ref{fig1}(b)], with a probe laser ( Rabi frequency $\Omega_p$) on $|1\rangle\leftrightarrow|2\rangle$ and a counterpropagating control laser (Rabi frequency $\Omega_c$) on $|2\rangle\leftrightarrow|3\rangle$. For an atom moving with velocity $v$ along the probe direction, the rotating-wave Hamiltonian is 
\begin{equation}
\begin{split}
H(v)=&-\hbar\Delta_p(v)\sigma_{22}-\hbar[\Delta_p(v)+\Delta_c(v)]\sigma_{33}\\
&-\frac{1}{2}\hbar\big(\Omega_p\sigma_{21}+\Omega_c\sigma_{32}+\mathrm{h.c.}\big),
\end{split}
\label{eq1}
\end{equation}
where $\sigma_{ij}=|i\rangle\langle j|$, and $\Delta_p(v)=\Delta_p+k_pv$, $\Delta_c(v)=\Delta_c-k_cv$ take into account both bare laser detuning ($\Delta_p,\Delta_c$) and Doppler detuning ($k_pv,-k_cv$). Taking into account the dissipation, the dynamics of the atom is described by the master equation, $\partial_t\rho(v)=-\frac{i} {\hbar}\big[H(v),\rho(v)\big]+\mathcal{L}[\rho]$,  where $\mathcal{L}[\rho]$ is the Lindblad operator that accounts for the decay processes in the atom. 

\textit{Kinematic saturation law in weak-probe limit.}---The transmission of the probe laser depends on the imaginary part $\alpha_3(v)$ of the density-matrix component $\rho_{21}(v)$. In the weak-probe limit, the steady state is obtained analytically as~\cite{GeaBanacloche1995}:
\begin{equation}
\alpha_3(v)=\frac{\gamma_{21}\,\Omega_p/2}{\gamma_{21}^2+\left[\Delta_p(v)-\dfrac{\Omega_c^2}{4[\Delta_p(v)+\Delta_c(v)]}\right]^2},
\label{eq2}
\end{equation}
where $\gamma_{21}=\Gamma_2/2$ with the decay rate $\Gamma_2$ from $|2\rangle$ to $|1\rangle$, and other dissipation channels are  reasonably neglected. Given the velocity distribution of thermal atoms, the Doppler averaged value of $\alpha_3(v)$ is: $\langle\alpha_{3}\rangle=\int_{-\infty}^{\infty}dv\,f(v)\,\alpha_3(v)$ with Maxwell-Boltzmann distribution $f(v)=1/(\sqrt{\pi}\,v_{0})\exp\!\left(-v^{2}/v_{0}^{2}\right)$. Here $v_0=\sqrt{2k_BT/M}$ is the most probable atomic velocity at temperature $T$ (with $M$ to be the mass of the atom). Propagating the probe through the vapor by Beer--Lambert law, the transmission of three-level system is given by:
\begin{equation}
\begin{split}
T_3&=\exp\left(-\frac{2k_pN\mu_{12}^2}{\hbar\varepsilon_0\Omega_p}\langle\alpha_{3}\rangle\right),
\end{split}
\label{eq3}
\end{equation}
where $\mu_{12}$ is the probe-transition dipole matrix element, $\varepsilon_0$ the permittivity of the vacuum, and  the atomic column density $N=nL$ with atomic density $n$ and vapor-cell length $L$. 



Under the VSDR condition, Doppler averaged value $\langle\alpha_{3}\rangle$ can be approximately written as [$v_0\gg\gamma_{21}/(Fk_p)$] [see Supplemental Material (SM) for details]:
\begin{equation}
    \langle\alpha_{3}\rangle\approx
\frac{\sqrt{\pi}\Omega_{p}F}{2k_{p}v_0}
\exp\left[-\left(\frac{v^{*}}{v_{0}}\right)^{2}\right],
\label{eq4}
\end{equation}
where $F=\gamma_{21}
/\sqrt{\gamma_{21}^{2}+k_{p}\Omega_{c}^{2}/(k_{c}-k_{p})}$ is the residual absorption of the dark state, i.e., the fraction of the bare absorption that survives EIT for the double-resonance class, and $\gamma_{21}/(Fk_p)$ is the narrow velocity window for participating in VSDR. Substituting Eq.~\eqref{eq4} into Eq.~\eqref{eq3} and using $\Delta T=T_3-T_2$, we can obtain the peak contrast $\Delta T_p$, where the two-level transmission is given by $T_2=T_3({\Omega_c=0})$.

Maximizing $\Delta T_p$ yields both the detuning corresponding to maximum peak ontrast and the saturation threshold. Under VSDR condition, only one quantity $v^*$($\Delta_c$ or $\Delta_p$) is independent. As a result, by solving $\partial\Delta T_p/\partial v^*=0$, one can find the maximum and the corresponding conditions. The optimum velocity satisfies (see SM for details):
\begin{equation}
    v^{m}=
\begin{cases}
0, & N\le N_{s},\\[6pt]
v_{0}\sqrt{\ln\!\left(\dfrac{N}{N_{s}}\right)}, & N>N_{s},
\end{cases}
\label{eq5}
\end{equation}
where the saturation threshold,
\begin{equation}
    N_{s}=\frac{v_{0}\hbar\varepsilon_{0}}{\sqrt{\pi}\,\mu_{12}^{2}}\,\frac{\ln F}{F-1}.
    \label{eq6}
\end{equation}
The detuning corresponding to maximum peak contrast can be obtained by $\Delta_p^m=-k_pv^m,\Delta_c^m=k_cv^m$. Below the threshold column density $N_s$, the only maximum is at $v^m=0$: the peak contrast stays on the EIT resonance and grows with $N$. Above $N_s$, the resonance point turns into a minimum of $\Delta T_p$ and the maximum moves to the velocity class $v_0\sqrt{\ln\!\left(N/N_s\right)}$. The enhanced off-resonant velocity-selective transmission observed in thermal Rydberg gases~\cite{Hu2025} being a particular instance of this mechanism. The saturation density $N_s$ is thus fixed by $F$ alone---the intermediate-state decoherence, the two wave numbers, and the control intensity---and diverges only logarithmically as $F\to0$, so saturation is unavoidable at any finite control intensity. Equation~\eqref{eq5} admits a transparent interpretation: at the optimum, the number of atoms in VSDR reaches its intrinsic optimum due to $Nf(v^m)=\frac{\hbar\varepsilon_{0}\ln F}{\pi\mu_{12}^{2}(F-1)}$ being constant, and the saturation value of the EIT peak contrast can be written as (see SM for details):
\begin{equation}
\Delta T_p^{\mathrm{sat}}=F^{\frac{F}{1-F}}-F^{\frac{1}{1-F}}.
\label{eq:dTmsat}
\end{equation}
The saturated peak contrast depends only on $F$---the fraction of the probe absorption that the control laser cannot remove from a double-resonant atom, i.e., the unavoidable absorption loss of Ref.~\cite{Meyer2021}. It is an intrinsic ceiling fixed solely by the competition between the intermediate-state decoherence and the control laser power: no adjustment of density or temperature can push the peak beyond it. Because only $F$ enters, the saturation law is universal: any vapor and any ladder transition sharing the same $F$ saturate at the same peak contrast and pinning, whatever the atom.

\begin{figure}[t]
\includegraphics[width=0.99\columnwidth]{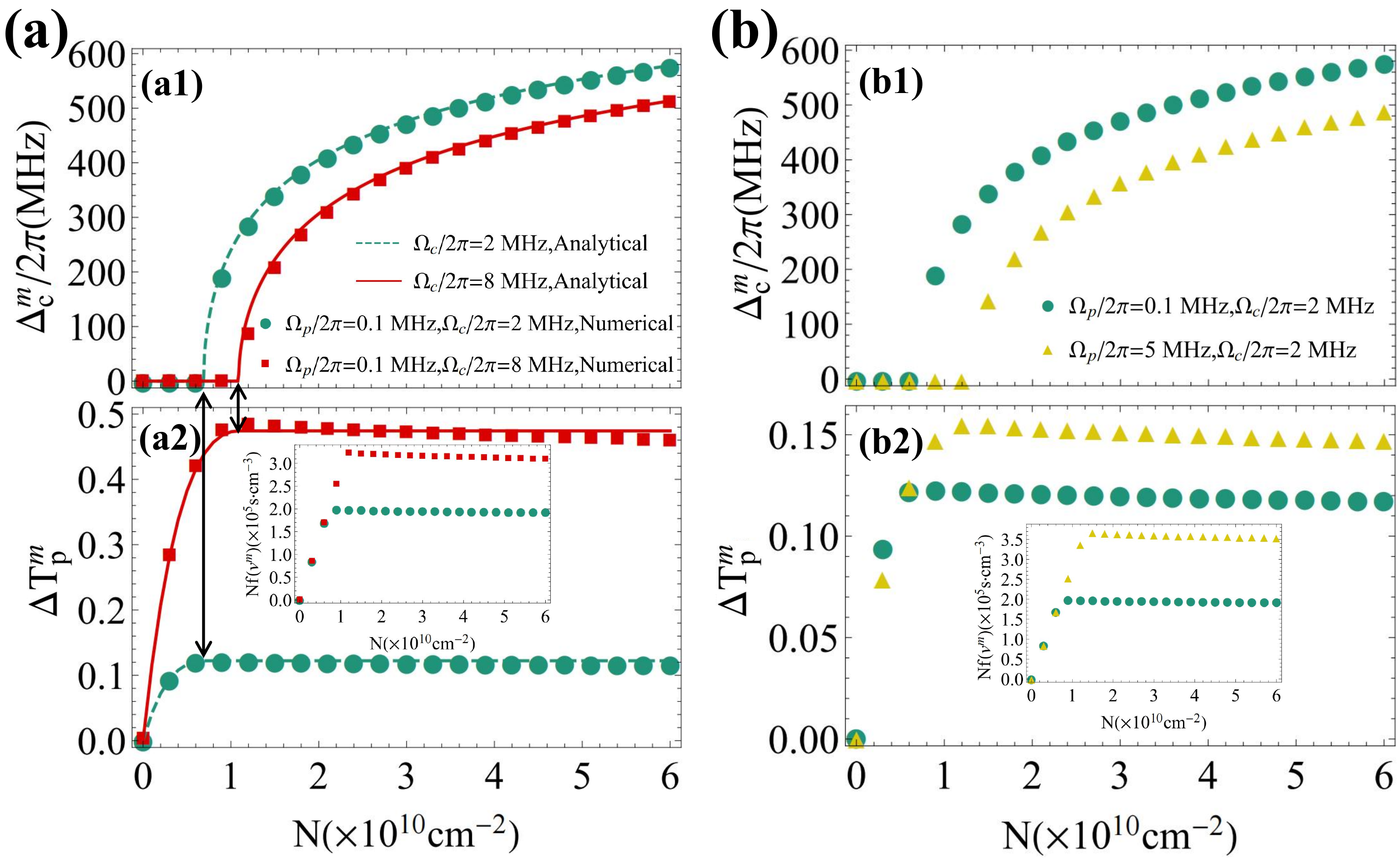}
\caption{\label{fig2}Saturation of the EIT peak in a room-temperature Cs vapor cell. (a) Weak-probe regime ($\Omega_p/2\pi=0.1$~MHz): (a1) optimal detuning $\Delta_c^m$ and (a2) maximum peak contrast $\Delta T_p^m$ versus atomic column density $N$ for $\Omega_c/2\pi=2$~MHz (green) and $8$~MHz (red); numerical solutions (symbols) agree quantitatively with the analytical results (curves). Arrows mark the saturation densities $N_s$. Inset of (a2): $Nf(v^m)$ changes with $N$. The slow decay of $\Delta T_p^m$ past $N_s$ is a effect of the Maxwell--Boltzmann asymmetry at $v^m$. (b) Beyond the weak-probe regime ($\Omega_p/2\pi=5$~MHz, triangles, compared with $0.1$~MHz, circles, at $\Omega_c/2\pi=2$~MHz): the same saturation behavior persists, while $N_s$ and $\Delta T_p^{\mathrm{sat}}$ acquires an additional dependence on the probe intensity.}
\end{figure}

We verify these results against numerical solutions of the master equation with Doppler averaging,  using the parameters of a room-temperature Cs vapor cell ($\lambda_p=852$~nm, $\lambda_c=510$~nm, $v_0\approx193$~m/s, $\Gamma_2/2\pi=5$ MHz). As shown in Fig.~\ref{fig2}(a), in the weak-probe regime ($\Omega_p/2\pi=0.1$~MHz) the numerics agree with the analytical threshold and detuning curve over the full scanned range of $N$ for both $\Omega_c/2\pi=2$ and $8$~MHz, and the saturated peak and $Nf(v^m)$ match quantitatively around the threshold; beyond it, both exhibit small deviations [Fig.~\ref{fig2}(a2) and inset]. The analytics result predicts that the saturated peak stays fixed as $N$ increases, while the numerics show that it decreases slightly and slowly once the optimum moves off resonance, and so does $Nf(v^m)$. Both deviations originate from the asymmetry of the Maxwell distribution about a nonzero selected velocity. As Eq.~(\ref{eq4}) shows, the analytics attributes the absorption to resonant atoms at exactly the selected velocity within a narrow velocity window, whereas in the numerics the absorption is carried by near-resonant atoms within the narrow velocity window. On EIT resonance the window straddles the symmetric center of the Maxwell distribution and the two descriptions coincide; off EIT resonance it straddles the asymmetric flank, so the participating column density effectively probed falls slightly below its pinned value---fewer atoms share the double resonance---and the deficit grows with the detuning. 

\textit{Beyond weak-probe limit.}---The saturation law above is derived in the weak-probe limit, but thermal-atom EIT experiments rarely operate there: the readout is limited by photon shot noise~\cite{kumar2017}, which recedes only with the square root of the probe power, so a weak probe means few detected photons and a noisy readout. To test whether the saturation law holds beyond this limit, we numerically solve the master equation with Doppler averaging at $\Omega_c/2\pi=2$~MHz, $\Omega_p/2\pi=5$~MHz. As shown in Fig.~\ref{fig2}(b), the peak contrast still saturates, its maximum still detunes, and the detuning still grows as $\sqrt{\ln(N/N_s)}$. While what differs is that the threshold 
$N_s$ and the value $\Delta T_p^{\mathrm{sat}}$ of the saturation law additionally depend on the probe intensity, with both increasing as the probe laser becomes stronger. The reason is power broadening: it widens the near double-resonance window and draws more atoms of the Maxwell distribution into near-double-resonance, so the participating column at the optimum---and with it the saturated peak---grows; the crossing of the two rates of Fig.~\ref{fig1}(c) shifts as well, and with it the threshold $N_s$. The saturation law is thus robust beyond the weak-probe limit: the probe power renormalizes its two parameters but leaves its mechanism untouched, namely the velocity-selected resource, the pinning, and the drift.

\textit{Sensitivity-optimal optical readout.}---The saturation law fixes where the maximum EIT peak sites and how high it rises. For any measurement built on EIT peak, the follow-up is where the sensitivity-optimal operating point locates. EIT-based sensing with optical readout, exemplified by Rydberg electrometry, is limited by photon shot noise~\cite{kumar2017,Yan2025,Ju2026}: with interferometric readout the technical noise of the probe laser is no longer the limit, and photon shot noise on the detector takes over as the sensitivity floor~\cite{kumar2017}, and state-of-the-art Rydberg electrometers operate essentially at it~\cite{Ju2026}. Further gains therefore hinge on optimizing photon shot noise. In Rydberg electrometry, a radio-frequency field Stark-shifts the Rydberg level, the shift is converted into a displacement $\delta$ of the two-photon resonance and read out through the EIT spectrum~\cite{Sedlacek2012,Jing2020}, and once the laser power and the vapor cell are fixed, the operating point on the spectrum is the remaining lever for sensitivity~\cite{Hu2025}. 

The EIT spectrum carries the field-dependent signal; the responsivity is the local spectral slope $|\kappa_0|=|dT_3/d\delta|$, and the photon-shot-noise-limited sensitivity, i.e., the minimum detectable displacement, is
\begin{equation}
\Delta\delta_{\min}=\frac{\sqrt{2\hbar\omega_pT_3}}{|\kappa_0|\sqrt{P_p}},
\label{eq:deltamin}
\end{equation}
where $P_p$ $(\propto\Omega_p^2)$ and $\omega_p$ are respectively the incident probe power and the frequency of the probe laser, and $T_3$ is the transmission at the operating point (the corresponding field sensitivity follows on dividing by the microwave responsivity $\partial\delta/\partial E$). Within a single EIT window the linewidth is fixed by the laser intensities, so a taller peak means a steeper slope while $T_3$ varies little: the slope dominates, and a larger $\Delta T_p$ is simply better. Across different VSDR windows, however, the two-level background $T_2$ and EIT contrast $\Delta T_p$ rise together, so that $T_3$ grows particularly steeply at far-detuned windows. $T_3$ penalty in Eq.~\eqref{eq:deltamin} then competes with the slope gain, and the sensitivity-optimal window is decided by the balance of the two factors for a fixed $P_p$. 

\begin{figure}[t]
\includegraphics[width=0.99\columnwidth]{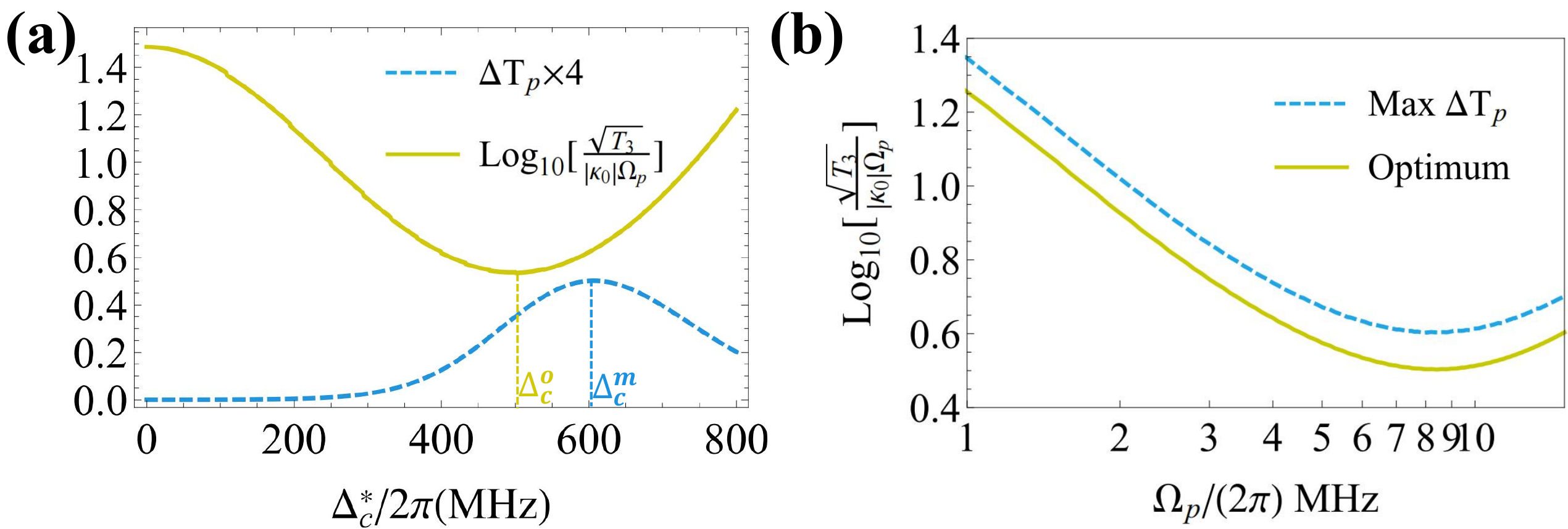}
\caption{\label{fig3}Optimal detuning for sensing differs from the peak maximum. (a) Transmission peak $\Delta T_p$ (dashed, magnified $\times4$) and the shot-noise figure of merit $\log_{10}[\sqrt{T_3}/(|\kappa_0|\Omega_p)]$ (solid), versus the velocity-selected detuning $\Delta_c^*$ for a room-temperature Cs Rydberg-EIT system ($\Omega_p/2\pi=6$~MHz, $\Omega_c/2\pi=1$~MHz); vertical lines mark $\Delta_c^m$ and $\Delta_c^o$, the detunings of the max-$\Delta T_p$ and optimal sensitivity, respectively. (b) The shot-noise figure of merit evaluated at $\Delta_c^m$ (dashed) and at $\Delta_c^o$ (solid) changes with the probe Rabi frequency ($\Omega_c/2\pi=1$~MHz).}
\end{figure}

To locate the optimum quantitatively, we evaluate
Eq.~\eqref{eq:deltamin} for a room-temperature Cs Rydberg-EIT system
under typical experimental conditions ($\Omega_p/2\pi=6$~MHz,
$\Omega_c/2\pi=1$~MHz, $N=15\times10^{10}$~cm$^{-2}$ with $L=5$~cm).
For each velocity-selected detuning $\Delta_c^*$ we minimize
$\Delta\delta_{\min}$ of Eq.~\eqref{eq:deltamin} within the
corresponding EIT window. Fig.~\ref{fig3}(a) shows the resulting figure of merit $\log_{10}[\sqrt{T_3}/(|\kappa_0|\Omega_p)]$, proportional to $\Delta\delta_{\min}$, versus $\Delta_c^*$. Two facts emerge from Fig.~\ref{fig3}(a). First, the window centered on the peak maximum $\Delta_c^m$ is indeed much more sensitive than the resonant window at $\Delta_c^*=0$~\cite{Hu2025}, which is the result under specific parameters. Second, the $\Delta_c^m$-centered window is not optimal: the figure of merit reaches its minimum at a detuning $\Delta_c^o\neq\Delta_c^m$, i.e., the sensitivity optimum does not coincide with the maximum $\Delta T_p$. The change in probe power has no effect on this ordering. As shown in Fig.~\ref{fig3}(b), the optimal sensitivity is always superior to that of the window centered at $\Delta_c^m$. The optimal sensitivity itself tracks the resonant-case evolution with the probe Rabi frequency~\cite{Ju2026}: it improves as the probe Rabi frequency increases until the power broadening effect competes with the slope, thereby determining a finite optimal probe power. The sensitivity at optimal $\Omega_p$ is 2 dB better than that of the window centered at $\Delta_c^m$.

\textit{Conclusion.}---In summary, the EIT transmission peak in a thermal vapor saturates because its resource is the velocity-selected double-resonance column density $N^*$, which has an optimal value. The saturation threshold $N_s$ and the saturated peak contrast $\Delta T_p^{\mathrm{sat}}$ are fixed by the intermediate-state decoherence, the two laser wavelengths and intensities. Beyond the threshold, the detuning corresponding to the saturated peak moves off resonance as $\sqrt{\ln(N/N_s)}$ so as to keep $N^*$ pinned at its optimal value, and the peak contrast decays slowly through the asymmetry of the thermal distribution. The same mechanism separates the operating points of maximal peak contrast and of optimal shot-noise-limited sensitivity. The conventional choice of maximizing the EIT peak therefore does not optimize the measurement. For vapor-cell sensing, where the lasers are tuned matters more than how many atoms fill the cell.

\begin{acknowledgments}
	We thank Z.-Y. Bai for useful discussion about calculation details. This work is supported by National Science Foundation of China under Grant No. 12274045, No. 12274046 and No.12347101.
\end{acknowledgments}
	
\bibliographystyle{apsrev4-2}
\bibliography{ref}
	
\end{document}